\documentclass[runningheads]{llncs}
\usepackage[T1]{fontenc}
\usepackage{graphicx}
\usepackage{hyperref}
\hypersetup{colorlinks=false, citecolor=blue}
\usepackage{amsmath}
\usepackage{amssymb,makeidx,enumerate}
\usepackage{multirow}
\usepackage{cite}
\usepackage{csquotes}

\begin{document}
\title{SimCPSR: Simple Contrastive Learning for Paper Submission Recommendation System}
%
%
\author{Duc H. Le\inst{1,2} \and
Tram T. Doan\inst{1,2} \and
Son T. Huynh\inst{1,2,3} \and Binh T. Nguyen\inst{1,2,3}\thanks{Corresponding Author: Binh T. Nguyen (ngtbinh@hcmus.edu.vn)}}
\authorrunning{Duc et al.}
%
\institute{University of Science, Ho Chi Minh City, Vietnam \and
Vietnam National University, Ho Chi Minh City, Vietnam \and
AISIA Research Lab, Ho Chi Minh City, Vietnam}

%
\titlerunning{Simple Contrastive Learning for Paper Submission Recommendation System}
\maketitle              
\begin{abstract}
The recommendation system plays a vital role in many areas, especially academic fields, to support researchers in submitting and increasing the acceptance of their work through the conference or journal selection process. This study proposes a transformer-based model using transfer learning as an efficient approach for the paper submission recommendation system. By combining essential information (such as the title, the abstract, and the list of keywords) with the aims \& scopes of journals, the model can recommend the Top K journals that maximize the acceptance of the paper. Our model had developed through two states: (i) Fine-tuning the pre-trained language model (LM) with a simple contrastive learning framework. We utilized a simple supervised contrastive objective to fine-tune all parameters, encouraging the LM to learn the document representation effectively. (ii) The fine-tuned LM was then trained on different combinations of the features for the downstream task. This study suggests a more advanced method for enhancing the efficiency of the paper submission recommendation system compared to previous approaches when we respectively achieve 0.5173, 0.8097, 0.8862, 0.9496 for Top 1, 3, 5, 10 accuracies on the test set for combining the title, abstract, and keywords as input features. Incorporating the journals' aims and scopes, our model shows an exciting result by getting 0.5194, 0.8112, 0.8866, 0.9496 respective to Top 1, 3, 5, and 10. We provide the implementation and datasets for further reference at \url{https://github.com/hduc-le/SimCPSR}.

\end{abstract}

\section{Introduction}

Recommendation systems have become more and more popular in almost all fields, and people are using these systems in different industries such as retail, media, news, streaming service, and e-commerce. See the benefit of recommendation systems in the development of the economy; many companies built recommendation systems that utilize historical data of customers to give them some relevant suggestions to meet customer satisfaction and improve the company's product. Various well-known recommendation systems include the recommendation systems of Spotify or Netflix in streaming services, Amazon in e-commerce branches, Google, Facebook, and Youtube in media. Besides, the recommendation system in academics has gained importance in recent years, including the paper submission system \cite{WANG20181,Son2020AnEA,Nguyen2021S2CFTAN,10.1007/978-3-030-79463-7_7}, the knowledge-based recommendation system \cite{article}, and paper suggestion \cite{8598708}.
Selecting a suitable journal for submitting new work is not easy for almost researchers, including young and experienced people. To support the researchers in choosing a relevant journal to increase their work acceptance opportunities, Wang and his coworkers proposed the recommendation system for computer science publications \cite{WANG20181} in the early stage and developed it by many other researchers later.

In this work, we aim to investigate the paper submission recommendation problem using general paper information and the aims and scopes of the journals as sufficient attributes in our method. Our target is to extract the semantic relationship between the paper submission and journal through those available features as well as possible. We expect the papers and journals' representations to be well-encoded, meaning close together for those that are semantically related and far away for contrast in the embedding space. We tackle the problem by applying the transformer architecture \cite{vanilla_Transformer} as an encoder to extract the input feature effectively and utilize the contrastive learning framework \cite{simclr, simcse} to enhance the model's robustness in the downstream task. The experimental results show that the proposed approach mostly outperforms the previous works. For example, in using the title, abstract, and keywords combined with journals' aims and scopes for training, our model gets 0.5194 for Top 1 accuracy and 0.9496 for Top 10 accuracy. 

The contribution of our work can be described as follows:
\begin{enumerate}[(a)]
    \item We propose a new framework for the paper submission recommendation problem using the transformer architecture, which shows a significant advance to tackle. The experimental results in Section \ref{experiment_sec} show that our approach has competitive performance compared to the previous works.
    \item Leveraging contrastive learning, the powerful method of sentence embedding, we enhance the framework's robustness in learning semantic relationships among documents or sentences.
    \item Our method provides a basic framework that can be extended further by applying other transformer models to achieve better performance in the paper submission recommendation problem.
\end{enumerate}

The paper can be organized as follows. First, Section \ref{relatedwork_sec} provides a brief overview of the approaches that tackled this problem using well-known tools and algorithms. Then, in Section \ref{method_sec}, we describe our main methods, and Section \ref{experiment_sec} shows the details of chosen experiments, including the training configurations, dataset, and achieved results. Finally, the paper ends with the conclusion and our discussion on future work.

\noindent

\section{Related work} \label{relatedwork_sec}
The idea of the paper recommendation system had developed by Wang, and his coworker in computer science publications \cite{WANG20181}. Wang used the Chi-square statistic and the term frequency-inverse document frequency (TF-IDF) for feature selection from the abstract of each paper submission and utilized linear logistic regression to classify relevant journals or conferences. The accuracy in his study is 61.37\% for the Top 3 recommended results. Son and colleagues later developed and proposed a new approach to improve the paper submission recommendation algorithm's performance using other additional features. The proposed method in \cite{Son2020AnEA} uses TF-IDF, the Chi-square statistic, and the one-hot encoding technique to extract parts from available information in each paper. They applied two machine learning models, namely Logistics Linear Regression (LLR) and Multi-layer Perceptrons (MLP), to the different combinations of features from the paper submission. Their proposed methods achieved outperformed results for the Top 3 accuracy, especially 89.07\% for the LLR model and 88.60\% for the MLP model when using a group of features, including the title, abstract, and keywords.

Regarding the problem, Dac et al. lately proposed a new approach \cite{Nguyen2021S2CFTAN} by applying two embedding methods, GloVe \cite{glove} and FastText \cite{fasttext}, combining to Convolutional Neural Network (CNN) \cite{cnn} and LSTM \cite{lstm} for feature extraction. They considered seven features groups: title, abstract, keywords, title + keyword, title + abstract, keyword + abstract, and title + keyword + abstract for training progress. The experimental results show that the combination of S2RSCS \cite{Son2020AnEA} and CNN + FastText, namely the proposed S2CFT \cite{Nguyen2021S2CFTAN} model has the best performance with the Top 1 accuracy is 68.11\% when using a mixture of attribute title + keyword + abstract, the accuracy at Top 3, 5, and 10 are 90.8\%, 96.25\%, and 99.21\% respectively.

Son and his coworkers continued to propose a new approach to the paper recommendation system. In their study \cite{10.1007/978-3-030-79463-7_7}, besides valuable papers' attributes, they used additional information from the \enquote{Aims and Scopes} of journals for input data. They collected a dataset containing 414512 articles and the corresponding aims and scopes in the journal of these papers. They built a new architecture that still uses FastText \cite{fasttext} for embedding, and the input data is available information of paper submission; they created a new feature by measuring the similarity between paper submission and the journals' aims \& scopes. Their proposed method is a practical approach to solving the paper submission recommendation problem.

In recent years, transformer architectures have succeeded in various fields of natural language processing (NLP) and computer vision (CV). One of the most significant works dedicated to that success is the Vanilla Transformer architecture \cite{vanilla_Transformer}. It is active through adopting the attention mechanism and differentially weighting the significance of each part of the input data. The popularity and efficiency of the transformer models are unquestionable; it contributes to carrying many other models taken off, the most famous being BERT (Bidirectional Encoder Representations from Transformers) \cite{bert}. BERT is pre-trained with large amounts of textual data and fine-tuned to achieve new state-of-the-art results in many NLP tasks such as semantic textual similarity (STS) or sentence classification. In addition, BERT has been considered a powerful embedding method for documents or sentences by sending those to the BERT layers and taking the average of the output layer or the output of the first token (the [CLS] token) to get the fixed-size embeddings.

Although BERT shows its impression on many NLP tasks, it exists some limitations in sentence embedding by standard approaches. Recently, the Contrastive Learning \cite{simclr} framework has become state-of-the-art in sentence embedding. Its idea conceptually describes a technique that aims to pull similar samples together and push dissimilar ones far away in the embedding space by the contrastive objective. One can apply contrastive learning to unlabeled and labeled data. As a result, fine-tuning transformer models by contrastive objective \cite{sbert, simcse, simclr} become an efficient method to perform better input representation, not only textual data but also image data \cite{simclr}.

\section{Methodology} \label{method_sec}
In this section, we present our approach for the paper submission recommendation problem and the evaluation metrics.

\subsection{Contrastive Learning} \label{cl_subsec}
Contrastive learning recently has become one of the popular frameworks that can be applied for supervised and unsupervised learning machine learning tasks. This technique aims to learn similar/dissimilar representations from data constructed from a set of paired samples $(x_i,x_i^+)$ semantically related effectively by pulling identical sample pairs together and pushing dissimilar ones apart in embedding space. 
The most significant promise of contrastive learning is utilizing a pre-trained language transformer model that was then fine-tuned with the contrastive objective to encode input into a good representation that can boost the performance of many downstream tasks.

We first leverage the supervised training dataset (as shown in Section \ref{data_desciption}) to construct a set of paired samples, $\displaystyle \mathcal{D} = \{(x_i, x_{i}^{+})\}_{i=1}^{m}$, for the contrastive fine-tuning process, in which we denote $x_{i}$ as the $i$'th sample that consists of the title, abstract, keywords and $x_{i}^{+}$ as semantically corresponding aims and scope. We follow the contrastive learning framework described in \cite{simcse}. For a mini-batch of $N$ pairs, let $\textbf{h}_{i}$ and $\textbf{h}_{i}^{+}$ denote the embedding or latent representation of $x_{i}$ and $x_{i}^{+}$; the contrastive objective was defined as:
\begin{equation} \label{ref_Eq1}
\ell_{i} = -\log \displaystyle\frac{e^{\textit{sim}(\textbf{h}_{i}, \textbf{h}_{i}^{+})/\tau}}{\sum_{i=1}^{N} e^{\textit{sim}(\textbf{h}_i, \textbf{h}_{j}^{+})/\tau}},
\end{equation}
where $\tau$ is a temperature hyper-parameter and 
$\textit{sim}\left(\mathbf{h}_1, \mathbf{h}_2\right)$ is the cosine similarity, which is:
\begin{equation*}
\textit{sim}\left(\mathbf{h}_1, \mathbf{h}_2\right) = \displaystyle \frac{\mathbf{h}_1^T \mathbf{h}_2}{\Vert \mathbf{h}_1\Vert \cdot \Vert \mathbf{h}_2\Vert}.
\end{equation*}

\subsection{Modeling} \label{modeling_sec}
In this study, we build a two-state model containing two consecutive procedures, conceptually fine-tuning the pre-trained LM with a simple contrastive learning framework as an encoder to encode each document or sentence into sentence embedding efficiently. Then, the fine-tuned LM can be applied to different combinations of the features for the downstream task to classify for the Top K accuracy on groups of attributes.

\begin{figure}
\includegraphics[width=\textwidth]{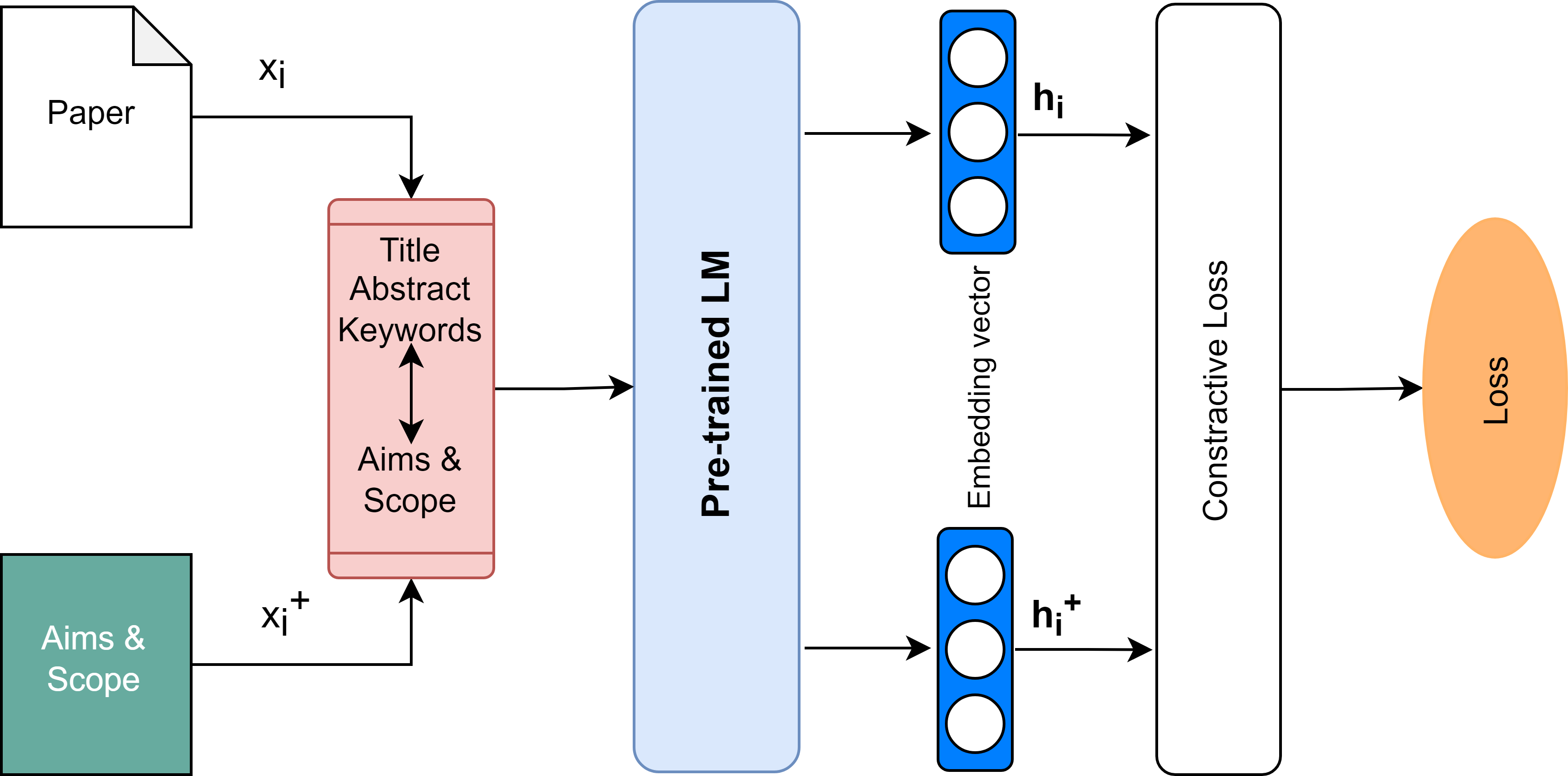}
\caption{Fine-tuning the pre-trained LM progress with a simple contrastive learning framework where $x_i$ and $x_{i}^{+}$ are considered two semantically related samples and those corresponding representation $\displaystyle \textbf{h}_i$, $\textbf{h}_{i}^{+}$.} \label{fig1}
\end{figure}

\textbf{Fine-tuning:} 
As mentioned in Section \ref{cl_subsec}, we consider the aims \& scope of the journal as a positive sample of the paper's title, abstract, and keywords, respectively. Finally, we perform fine-tuning the pre-trained Distil-RoBERTa (a distilled version of the pre-trained RoBERTa \cite{roberta}) via the contrastive objective (Eq. \ref{ref_Eq1}) on the set of paired samples to fine-tune all parameters as depicted in Figure \ref{fig1}.

\textbf{Downstream task:} 
We consider the fine-tuned LM as a backbone for the classification task. Therefore, we train it on different combinations of the features, either using available attributes of each paper submission or combining the paper information and the journals' aims. In this section, we describe two different use-cases in our experiments.

\begin{figure}
\includegraphics[width=\textwidth]{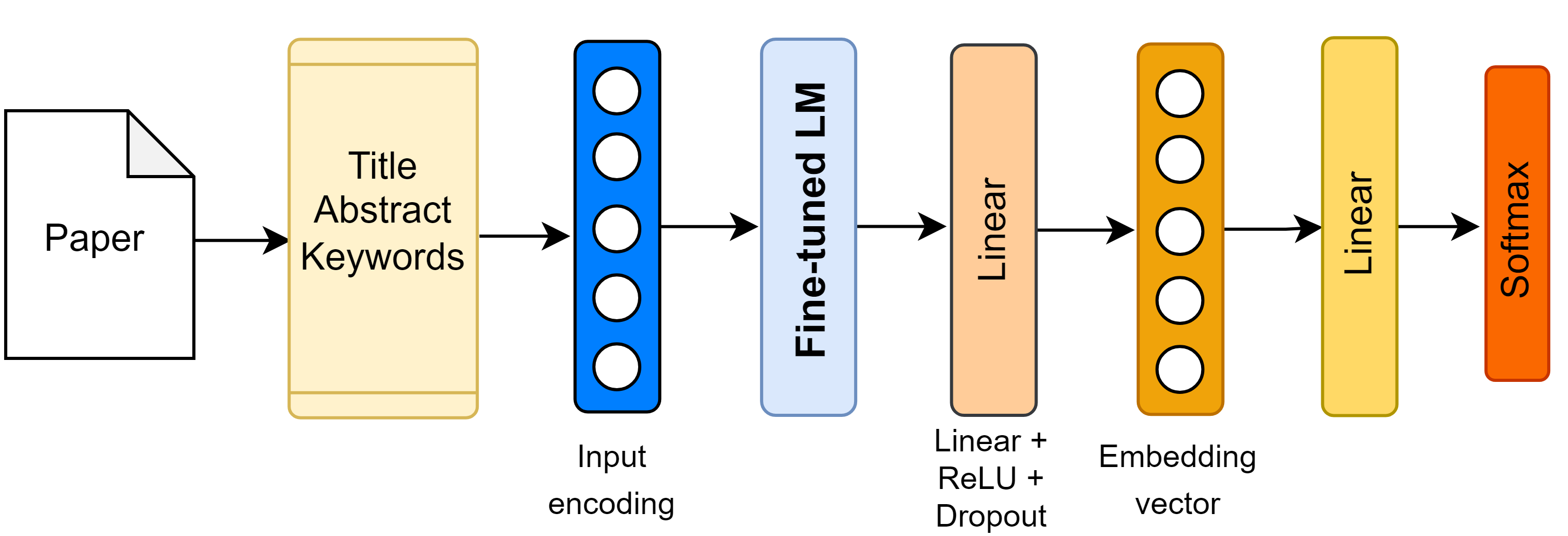}
\caption{The architecture of the fine-tuned LM using paper information.} \label{fig2}
\end{figure}

\textbf{Models using paper information} 
For this case, we extract helpful information from each paper submission, including Title (T), Abstract(A), and Keywords (K). This information is combined into seven different combinations of attributes: Title(T), Abstract(A), Keywords(K), Title + Abstract (TA), Title + Keywords (TK), Abstract + Keywords (AK), Title + Abstract + Keywords (TAK). The fine-tuned LM plays as the encoder to encode the batch of inputs into 768-dimensional embeddings, which are further forward propagated through an additional linear layer with ReLU activation and dropout to avoid overfitting. The last linear layer with Softmax activation acts as a classifier to output the N-probabilities that the given paper could belong to the respective journal. To identify the K-relevant journals that maximize the probability of acceptance, we choose the Top K of maximum values. The illustration for this case is in Figure \ref{fig2}.

\textbf{Models using paper information and Journals' Aims \& Scope}
Besides the available attributes on paper, we use Journals' Aims \& Scopes as potential external features. We end up with seven new combinations of features for the input data, Title + Aims and Scopes (TS), Abstract + Aims and Scopes (AS), Keywords + Aims and Scopes (KS), Title + Abstract + Aims and Scopes (TAS), Title +Keywords + Aims and Scopes (TKS), Abstract + Keywords + Aims and Scopes (AKS), Title + Abstract + Keywords + Aims and Scopes (TAKS). 

\begin{figure}
\includegraphics[width=\textwidth]{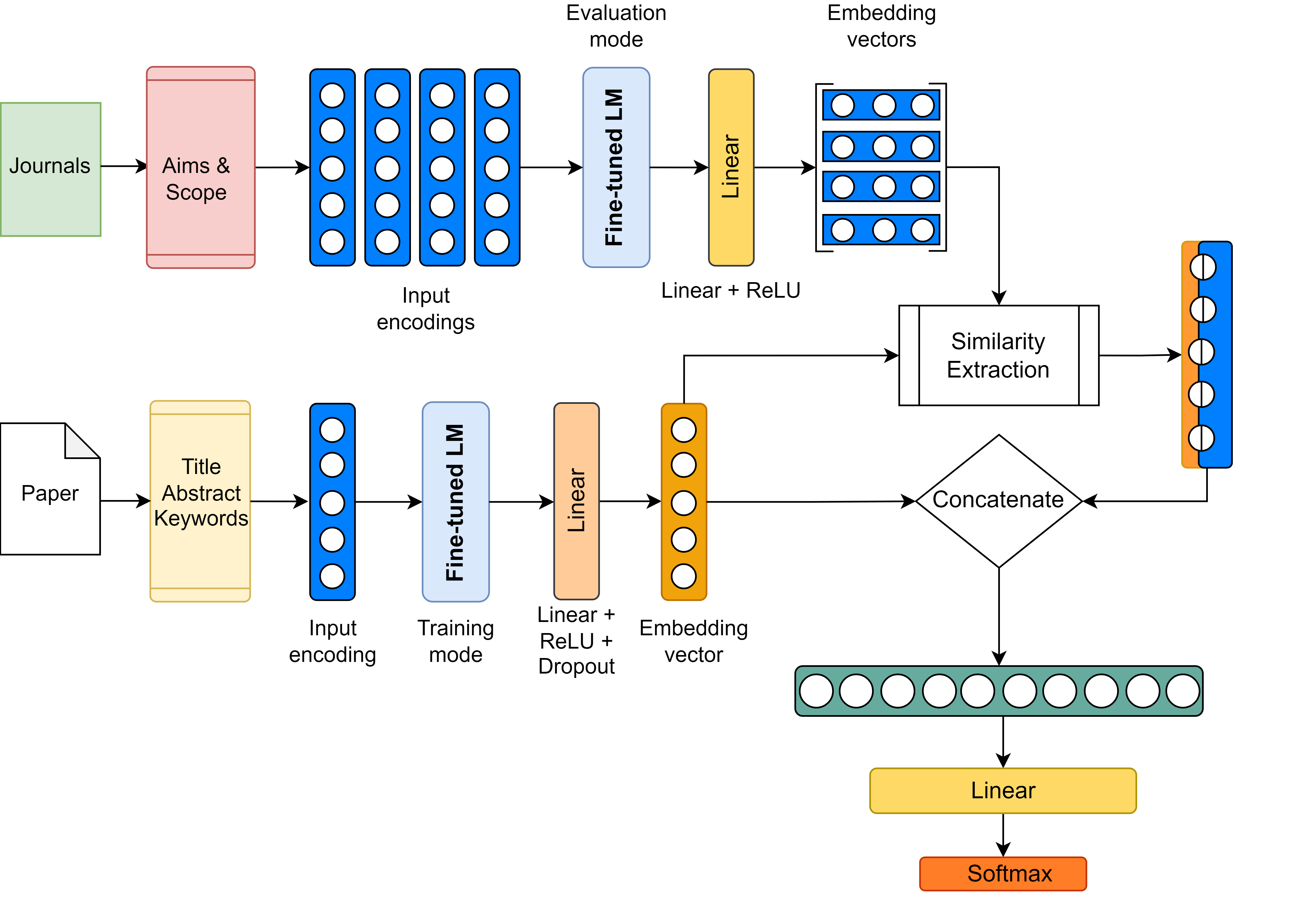}
\caption{The architecture of the fine-tuned LM using paper information and journal's aims.} \label{fig3}
\end{figure}

Conceptually, we define two sub-branches and the main branch of this architecture. In the first sub-branch, we reuse the fine-tuned LM to encode the external feature into the embeddings and pass it through a linear layer with ReLU activation for dimensional reduction. The second sub-branch is the same as the model using paper information described previously. Remarkably, the paper feature, output from the second sub-branch, will contribute to the following steps. First, we extract the similarity between it and external features produced from the first sub-branch using the cosine similarity and then concatenate it with these cosine features. Then, in the main branch, we feed that joined information to a linear layer and softmax activation to compute the probability of paper submission belonging to the journals and sort them in descending order to return the top list of recommended items. The illustration for the architecture can be found in Figure \ref{fig3}.

\subsection{Evaluation metrics} \label{metric_sec}

As the previous approach \cite{10.1007/978-3-030-79463-7_7}, Son et al. used Accuracy@K to evaluate the performance of the proposed model, where K = 1, 3, 5, 10. The Accuracy@K is the ratio between the number of correct items at each K and the K number of recommended items, described in the formula follows:
\begin{equation}
    P_{Top-K} =\frac{\text{The number of relevant items}}{\text{The number of viewed items}}
\end{equation}

\section{Experiments} \label{experiment_sec}
This section will present how we conducted the experiments for the proposed methods and compare them with the previous approaches for the main problem. 

\subsection{Experimental settings}
Our experiments are run on Google Colab Pro, mostly with Tesla P100-PCIE 16GB VRAM GPU accelerator, and implemented in the PyTorch framework \cite{pytorch}. In addition, we utilize the HuggingFace library \cite{huggingface_transformers}, one of the most popular NLP frameworks that provide APIs to download and use pre-trained transformer models easily. 

\subsection{Datasets}\label{data_desciption}
Our experiments use the same dataset and preprocessing techniques as used in Son et al. 's work \cite{10.1007/978-3-030-79463-7_7}, including 414512 papers, where there are 331464 papers used for training and 83048 ones used for testing. These consist of paper submissions' information (the title, abstract, and list of keywords) and their labels. Besides, 351 aims and scopes belong to the journals play as external contributed features.

\textbf{Data preprocessing}
Excluding the size of the dataset, the quality of the preprocessed data affects the model's performance very essentially, so data preprocessing is a crucial step in almost machine learning tasks; this process is more critical in NLP problems. The preprocessing progress includes some steps such as (1) lowercase text, (2) removing punctuations, (3) removing URLs, (4) removing stop words downloaded from the Natural Language Toolkit (NLTK6), and some other ones we define, (5) removing unnecessary spacing,(6) removing not-be-alphabet text.
In our work, we apply data preprocessing techniques for two kinds of data, including general paper information and the aims and scopes of the journal.
\subsection{Training details}
We start by using a pre-trained Distil-RoBERTa model and fine-tune it on the set of paired samples where each instance is constructed by pairing the combination of the title, abstract, and keywords with the corresponding aims \& scope of the journal to take the [CLS] representation as to the sentence embedding. Lastly, we fine-tune the model's parameters using the contrastive objective (Eq. \ref{ref_Eq1}) to lead it more robust in extracting the semantic relationship among sentences. During the experiment, we found that the AdamW \cite{adamW_paper} optimizer, a variant of the optimizer Adam \cite{adam} uses weight decay to avoid overfitting, is the efficient choice for deep networks like our architecture.

According to \ref{modeling_sec}, we solve the classification problem as the downstream task in which we train the fine-tuned Distil-RoBERTa model on different combinations of features and use the AdamW to optimize Cross-Entropy loss. Finally, we put the Softmax layer to archive the Top K values representing the probability of given inputs belonging to corresponding labels and compute the accuracy at each K value as described in Section \ref{metric_sec}

\subsection{Results}

\begin{figure}
\includegraphics[width=\textwidth]{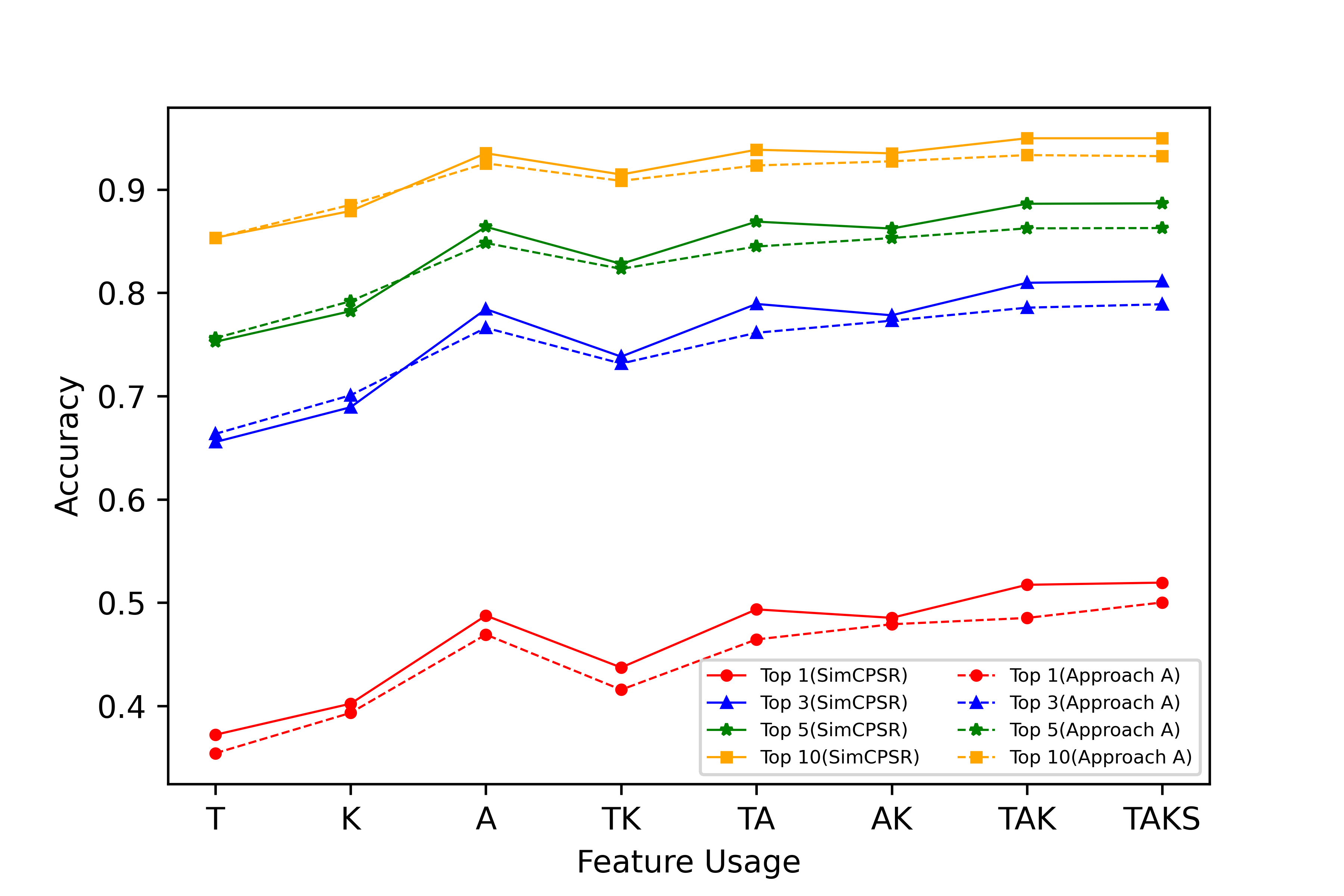}
\caption{The performance between the proposed SimCPSR approach and Approach A for different features.} \label{fig4}
\end{figure}

Our experiments are done on different combinations of attributes using two models defined previously: \textbf{Models using paper information} and \textbf{Models using paper information and Journals' Aims \& Scope}. We compare the performance of our proposed model with the approach \cite{10.1007/978-3-030-79463-7_7} (namely Approach A), the experimental results described in Table \ref{table1} and Figure \ref{fig4}. 

Firstly, for models only using paper information, the results show that the proposed method performs better than Approach A for combinations of attributes such as A, TK, TA, AK, and TAK. Especially, our approach has the best outcome for the group of features TAK in Accuracy@K as 0.5173, 0.8097, 0.8862, and 0.9496, where K = 1, 3, 5, and 10. Meanwhile, Approach A with the same input has 0.4852, 0.7856, 0.8624, and 0.9333, respectively. However, the proposed model's performance is lower than Approach A for the title (T) and keywords (K) inputs at the Top 3 and 5 in the accuracy; this difference is slight. Besides, the accuracy of the proposed model at the Top 1 surpasses the previous method for the title and keywords (TK) input; specifically, the Accuracy@K (with K = 1) of the proposed model is 0.3721 and 0.4022; meanwhile, the performance of Approach A is 0.3542, 0.3933, respectively. In addition, compared to the accuracy of Approach A in the Top 10, our approach gives better performance for the title (T) feature, and it has a little lower outcome for the keywords (K) attribute.

Secondly, for models using paper information and Journals' Aims \& Scope, the proposed method's performance surpasses Approach A when using types of input data such as AS, TAS, AKS, and TAKS. For example, the best performance of the proposed approach in Accuracy@K (K=1, 3, 5, 10) is 0.5194, 0.8112, 0.8866, and 0.9496 when using all features (TAKS), while the performance of Approach A is lower than, which are 0.5002, 0.7889, 0.8627, and 0.9323 respectively. Although for the remaining input groups (namely TS, KS, and TKS), the Accuracy@K (K=1, 3, 5, 10) of the proposed model is lower than Approach A, excluding only one case Accuracy@10 with TKS as input data, the accuracy of the proposed approach is slightly greater than Approach A.

Finally, one can see that both the proposed method and approach A can help to improve performance when using additional information \enquote{Aims \& Scopes} of the journals. Using Aims \& Scopes' results in our proposed method is better than not using one. Except in the cases of using types of features such as title (T), the keyword (K), title, and keyword (TK), using the additional attribute \enquote{Aims \& Scopes} is not helpful to increase performance. For example, the best performance of the proposed method in the Accuracy@K (K=1, 3, 5, 10) is 0.5194, 0.8112, 0.8866, and 0.9496 when using combinations of attributes TAKS; meanwhile, the outcome when not using the aims of the journals are 0.5173, 0.8097, 0.8862 and 0.9496 respectively.

In summary, our proposed approach, SimCPSR, has outperformed performance compared to the previous method; this can be considered a new approach for the paper submission recommendation system. Remarkably, using additional information \enquote{Aims \& Scopes} can help improve models' efficiency. Besides, we can see the importance of the abstract feature; all models using input data containing this feature have better performance than the models using input without abstract information. Therefore, the abstract factor is essential for the paper submission recommendation problem.

\begin{table}[]
\begin{center}
\caption{The performance of the proposed models compared to Son et al.'s results in \cite{10.1007/978-3-030-79463-7_7}}

\begin{tabular}{clllll}
\hline
\multicolumn{1}{|c|}{\textbf{Method}} & \multicolumn{1}{c|}{\textbf{Feature Usage}} & \multicolumn{1}{c|}{\textbf{Top 1}} & \multicolumn{1}{c|}{\textbf{Top 3}} & \multicolumn{1}{c|}{\textbf{Top 5}} & \multicolumn{1}{c|}{\textbf{Top 10}} \\ \hline
\multicolumn{1}{|c|}{\multirow{14}{*}{Approach A}} & \multicolumn{1}{l|}{T} & \multicolumn{1}{l|}{0.3542} & \multicolumn{1}{l|}{\textbf{0.6634}} & \multicolumn{1}{l|}{\textbf{0.7561}} & \multicolumn{1}{l|}{0.8532} \\ \cline{2-6} 
\multicolumn{1}{|c|}{} & \multicolumn{1}{l|}{TS} & \multicolumn{1}{l|}{\textbf{0.4015}} & \multicolumn{1}{l|}{\textbf{0.6991}} & \multicolumn{1}{l|}{\textbf{0.7971}} & \multicolumn{1}{l|}{0.8951} \\ \cline{2-6} 
\multicolumn{1}{|c|}{} & \multicolumn{1}{l|}{K} & \multicolumn{1}{l|}{0.3933} & \multicolumn{1}{l|}{\textbf{0.7008}} & \multicolumn{1}{l|}{\textbf{0.7919}} & \multicolumn{1}{l|}{\textbf{0.8852}} \\ \cline{2-6} 
\multicolumn{1}{|c|}{} & \multicolumn{1}{l|}{KS} & \multicolumn{1}{l|}{\textbf{0.4284}} & \multicolumn{1}{l|}{\textbf{0.7256}} & \multicolumn{1}{l|}{\textbf{0.8189}} & \multicolumn{1}{l|}{\textbf{0.9075}} \\ \cline{2-6} 
\multicolumn{1}{|c|}{} & \multicolumn{1}{l|}{A} & \multicolumn{1}{l|}{0.4691} & \multicolumn{1}{l|}{0.7661} & \multicolumn{1}{l|}{0.8482} & \multicolumn{1}{l|}{0.9253} \\ \cline{2-6} 
\multicolumn{1}{|c|}{} & \multicolumn{1}{l|}{AS} & \multicolumn{1}{l|}{0.477} & \multicolumn{1}{l|}{0.7662} & \multicolumn{1}{l|}{0.8488} & \multicolumn{1}{l|}{0.9258} \\ \cline{2-6} 
\multicolumn{1}{|c|}{} & \multicolumn{1}{l|}{TK} & \multicolumn{1}{l|}{0.4157} & \multicolumn{1}{l|}{0.7315} & \multicolumn{1}{l|}{0.8232} & \multicolumn{1}{l|}{0.9084} \\ \cline{2-6} 
\multicolumn{1}{|c|}{} & \multicolumn{1}{l|}{TKS} & \multicolumn{1}{l|}{\textbf{0.4475}} & \multicolumn{1}{l|}{\textbf{0.7490}} & \multicolumn{1}{l|}{\textbf{0.8302}} & \multicolumn{1}{l|}{0.9127} \\ \cline{2-6} 
\multicolumn{1}{|c|}{} & \multicolumn{1}{l|}{TA} & \multicolumn{1}{l|}{0.4644} & \multicolumn{1}{l|}{0.7613} & \multicolumn{1}{l|}{0.8448} & \multicolumn{1}{l|}{0.9233} \\ \cline{2-6} 
\multicolumn{1}{|c|}{} & \multicolumn{1}{l|}{TAS} & \multicolumn{1}{l|}{0.4828} & \multicolumn{1}{l|}{0.7754} & \multicolumn{1}{l|}{0.8536} & \multicolumn{1}{l|}{0.9276} \\ \cline{2-6} 
\multicolumn{1}{|c|}{} & \multicolumn{1}{l|}{AK} & \multicolumn{1}{l|}{0.4791} & \multicolumn{1}{l|}{0.7730} & \multicolumn{1}{l|}{0.8530} & \multicolumn{1}{l|}{0.9273} \\ \cline{2-6} 
\multicolumn{1}{|c|}{} & \multicolumn{1}{l|}{AKS} & \multicolumn{1}{l|}{0.4951} & \multicolumn{1}{l|}{0.7830} & \multicolumn{1}{l|}{0.8602} & \multicolumn{1}{l|}{0.9304} \\ \cline{2-6} 
\multicolumn{1}{|c|}{} & \multicolumn{1}{l|}{TAK} & \multicolumn{1}{l|}{0.4852} & \multicolumn{1}{l|}{0.7856} & \multicolumn{1}{l|}{0.8624} & \multicolumn{1}{l|}{0.9333} \\ \cline{2-6} 
\multicolumn{1}{|c|}{} & \multicolumn{1}{l|}{TAKS} & \multicolumn{1}{l|}{0.5002} & \multicolumn{1}{l|}{0.7889} & \multicolumn{1}{l|}{0.8627} & \multicolumn{1}{l|}{0.9323} \\ \hline
\multicolumn{6}{c}{} \\ \hline
\multicolumn{1}{|c|}{\multirow{14}{*}{SimCPSR}} & \multicolumn{1}{l|}{T} & \multicolumn{1}{l|}{\textbf{0.3721}} & \multicolumn{1}{l|}{0.6555} & \multicolumn{1}{l|}{0.7526} & \multicolumn{1}{l|}{\textbf{0.8533}} \\ \cline{2-6} 
\multicolumn{1}{|c|}{} & \multicolumn{1}{l|}{TS} & \multicolumn{1}{l|}{0.3737} & \multicolumn{1}{l|}{0.6553} & \multicolumn{1}{l|}{0.7513} & \multicolumn{1}{l|}{0.8523} \\ \cline{2-6} 
\multicolumn{1}{|c|}{} & \multicolumn{1}{l|}{K} & \multicolumn{1}{l|}{\textbf{0.4022}} & \multicolumn{1}{l|}{0.6892} & \multicolumn{1}{l|}{0.7822} & \multicolumn{1}{l|}{0.8792} \\ \cline{2-6} 
\multicolumn{1}{|c|}{} & \multicolumn{1}{l|}{KS} & \multicolumn{1}{l|}{0.4015} & \multicolumn{1}{l|}{0.6921} & \multicolumn{1}{l|}{0.7839} & \multicolumn{1}{l|}{0.8784} \\ \cline{2-6} 
\multicolumn{1}{|c|}{} & \multicolumn{1}{l|}{A} & \multicolumn{1}{l|}{\textbf{0.4875}} & \multicolumn{1}{l|}{\textbf{0.7842}} & \multicolumn{1}{l|}{\textbf{0.8639}} & \multicolumn{1}{l|}{\textbf{0.9351}} \\ \cline{2-6} 
\multicolumn{1}{|c|}{} & \multicolumn{1}{l|}{AS} & \multicolumn{1}{l|}{\textbf{0.4886}} & \multicolumn{1}{l|}{\textbf{0.7849}} & \multicolumn{1}{l|}{\textbf{0.8642}} & \multicolumn{1}{l|}{\textbf{0.9353}} \\ \cline{2-6} 
\multicolumn{1}{|c|}{} & \multicolumn{1}{l|}{TK} & \multicolumn{1}{l|}{\textbf{0.4372}} & \multicolumn{1}{l|}{\textbf{0.7382}} & \multicolumn{1}{l|}{\textbf{0.8280}} & \multicolumn{1}{l|}{\textbf{0.9145}} \\ \cline{2-6} 
\multicolumn{1}{|c|}{} & \multicolumn{1}{l|}{TKS} & \multicolumn{1}{l|}{0.4367} & \multicolumn{1}{l|}{0.7354} & \multicolumn{1}{l|}{0.8268} & \multicolumn{1}{l|}{\textbf{0.9128}} \\ \cline{2-6} 
\multicolumn{1}{|c|}{} & \multicolumn{1}{l|}{TA} & \multicolumn{1}{l|}{\textbf{0.4935}} & \multicolumn{1}{l|}{\textbf{0.7892}} & \multicolumn{1}{l|}{\textbf{0.8689}} & \multicolumn{1}{l|}{\textbf{0.9385}} \\ \cline{2-6} 
\multicolumn{1}{|c|}{} & \multicolumn{1}{l|}{TAS} & \multicolumn{1}{l|}{\textbf{0.5014}} & \multicolumn{1}{l|}{\textbf{0.7920}} & \multicolumn{1}{l|}{\textbf{0.8729}} & \multicolumn{1}{l|}{\textbf{0.9428}} \\ \cline{2-6} 
\multicolumn{1}{|c|}{} & \multicolumn{1}{l|}{AK} & \multicolumn{1}{l|}{\textbf{0.4853}} & \multicolumn{1}{l|}{\textbf{0.7782}} & \multicolumn{1}{l|}{\textbf{0.8622}} & \multicolumn{1}{l|}{\textbf{0.9350}} \\ \cline{2-6} 
\multicolumn{1}{|c|}{} & \multicolumn{1}{l|}{AKS} & \multicolumn{1}{l|}{\textbf{0.5030}} & \multicolumn{1}{l|}{\textbf{0.7964}} & \multicolumn{1}{l|}{\textbf{0.8765}} & \multicolumn{1}{l|}{\textbf{0.9435}} \\ \cline{2-6} 
\multicolumn{1}{|c|}{} & \multicolumn{1}{l|}{TAK} & \multicolumn{1}{l|}{\textbf{0.5173}} & \multicolumn{1}{l|}{\textbf{0.8097}} & \multicolumn{1}{l|}{\textbf{0.8862}} & \multicolumn{1}{l|}{\textbf{0.9496}} \\ \cline{2-6} 
\multicolumn{1}{|c|}{} & \multicolumn{1}{l|}{TAKS} & \multicolumn{1}{l|}{\textbf{0.5194}} & \multicolumn{1}{l|}{\textbf{0.8112}} & \multicolumn{1}{l|}{\textbf{0.8866}} & \multicolumn{1}{l|}{\textbf{0.9496}} \\ \hline
\end{tabular}
\label{table1}
\end{center}
\end{table}

\section{Conclusion and Further Works} \label{conclusion_sec}
We presented a transformer-based approach to the paper submission recommendation system with simple contrastive learning. The proposed method utilized a simple supervised contrastive objective to fine-tune all parameters in the pre-trained LM for embedding input data and the fine-tuned LM to train different combinations of the features using available paper information and the journal's aims for downstream task. The experimental results show that the proposed approach has competitive performance and is an advanced method for enhancing the efficiency of the paper submission recommendation system. Furthermore, we will continue improving the proposed algorithm's performance and apply them to different datasets that belong to various areas in the future. 

\section*{Acknowledgement}
Son Huynh Thanh was funded by Vingroup JSC and supported by the Master, Ph.D. Scholarship Programme of Vingroup Innovation Foundation (VINIF), Institute of Big Data, code VINIF.2021.ThS.18.

%
%
%

\bibliographystyle{splncs04}
\bibliography{bibliography.bib}

\end{document}